\documentclass[12pt]{article}
\usepackage{amsmath,amssymb,mathrsfs,hyperref,slashed,bbm,xcolor}
\newcommand{\bra}[1]{\ensuremath{\left\langle#1\right|}}
\newcommand{\ket}[1]{\ensuremath{\left|#1\right\rangle}}
\def\({\left(} \def\){\right)}

\begin{document}
\title{\vspace{-1.8in}
{Requiem for the ideal clock}}
\author{\large K.L.H. Bryan${}^{(1)}$,  A.J.M. Medved${}^{(1,2)}$
\\
\vspace{-.5in} \hspace{-1.5in} \vbox{
 \begin{flushleft}
$^{\textrm{\normalsize (1)\ Department of Physics \& Electronics, Rhodes University,
  Grahamstown 6140, South Africa}}$
$^{\textrm{\normalsize (2)\ National Institute for Theoretical Physics (NITheP), Western Cape 7602,
South Africa}}$
\\ \small \hspace{1.07in}
  g08b1231@ru.ac.za,\  j.medved@ru.ac.za
\end{flushleft}
}}
\date{}
\maketitle

\begin{abstract}
 
The ‘problem of time’ remains an unresolved issue in all known physical descriptions of the Universe. One aspect of this problem is the conspicuous absence of time in the Wheeler-Dewitt equation, which is the analogue of the Schrodinger equation for the Universal wavefunction. Page and Wootters famously addressed this problem by providing a mechanism for effectively introducing time evolution into this timeless cosmological picture. Their method, which is sometimes called the conditional probability interpretation (CPI), requires the identification of an internal clock system that is meant to keep time for the remainder of the Universe. Most investigations into this idea employ the idealized limit of a non-interacting clock system, the so-called ideal clock. However, by allowing for interactions, we have found the counter-intuitive result that a non-interacting clock is not necessarily the optimal choice, even if it is ‘ideal’. In particular, the uncertainty that is associated with the physical measurement of an atomic clock is found to decrease monotonically as the interactions grow stronger. This observation, which is reinforced by a previous study using a semi-classical clock, paves the way to an independent argument that is based on the energy conservation of any isolated system. Our conclusion is that ideal clocks must be prohibited from the CPI when recovering cosmological time evolution. Interactions are necessary for describing time evolution as a strict matter of principle. Lastly, we also consider the implications of this result for the experience of time in the evolution of the Universe.

\end{abstract}

\newpage
\section{Introduction}
\label{1}
\subsection{Background and Motivations}
\label{1.1}    
The `problem of time' and its various components have been described {\em ad nauseum} within the vast collection of literature on the subject.
    As the second law of thermodynamics is a rare physical principle providing a direction from the past to the future,
    it is usually the first facet of such discussions but is closely followed by
    a second. This would be the lack of not only a direction but a common description of time in the two fundamental theories --- quantum theory and general relativity --- which respectively view time as an external parameter and an abstract spacetime dimension.
This is  an important disagreement to settle before one
  even  contemplates broaching the daunting subject of quantum gravity. But, even with these issues aside,
    a subtle point is often excluded from such discussions: There is a distinction between the usual parametrizations of time as it appears in mathematical expressions and
the emergent phenomenon of time evolution as it is understood through our life-long experiences.
    Most of our trouble in describing time actually lies within the purview of the latter. Whether it is the absolute time of the quantum world
    or the abstract  dimension of classical relativity, our restricted movement in time remains an unresolved puzzle.
   
    This lack of an explanation for our passage through time is brought to the fore by a third facet of the
    problem: the timelessness of the Universe.
    Wheeler and De Witt introduced this notion in the form of
a  mathematical statement \cite{dewitt}, since named after them, 
    \begin{equation}
     \label{wdw}
     \hat{H}\ket{\psi}\;=\;0\;.
    \end{equation}
    Here, $\hat{H}$ represents a quantum analogue 
  of the Hamiltonian constraint of general relativity (although the equation should really be viewed as  semiclassical)  and $\ket{\psi}$ represents
    the total state of the Universe. This somewhat {\em ad hoc} but generally accepted equation enforces a total energy of zero
    for the Universe and, as a consequence, imposes an entirely static description on $\ket{\psi}$.
  Yet,  even if the mathematics is sound, the imposed timelessness 
on the state of the Universe is at odds with our experiences from
    within. The question is then the same as before, only more so: Why do we experience a
    directed evolution in time?

This paradoxical situation was taken up by Page and Wootters, who managed to resolve
    it into a workable theory which is indeed capable of describing evolution \cite{PW} (see also \cite{aha}).
    The premise is to divide the entire state $\ket{\psi}$ into two strongly entangled subsystems: a  `clock'  $C$ and the remainder
of the Universe $R$. (The entanglement is necessarily quantum.)
    The evolution of $R$ is then to be described in terms
    of a measurement of one of the clock's variables. To further clarify, at no point is time measured directly, as there is no appearance of time in the conventional sense.
    Rather, an eigenvalue of $C$, such as the location of its  center of mass $\vec{x}_{CM}$, is used to provide an effective time variable. The
    evolution of $\vec{x}_{CM}$ would then be accessible to $R$ because of
its mutual entanglement with $C$. A more detailed description is provided
in Appendix~\ref{PWM}.
   
The Page--Wootters' approach has met with some amount of resistance; most notably, Kuchar's concerns  that their clock  could not
    describe a succession of time measurements and, therefore, no description of evolution would be possible \cite{kuchar}.
    These concerns have since been been countered by Dolby \cite{dolby} (and independently by Giovannetti {\em et al.} \cite{lloyd}, also see 
\cite{add1,add2}), who furthered the
   the Page--Wootters' treatment  while renaming it as the conditional
probability interpretation (CPI).~\footnote{For a contrary opinion regarding Dolby`s 
resolution, see \cite{rovellietal}.}  
  Dolby showed that the CPI is consistent by adopting  an integration  variable to `sync' $C$ and $R$ and thus play the role of an abstract time parameter. 
  This integration is basically  the same  as tracing out the clock system, a
    procedure which is favored by many others.

    There is, however, another concern which presents a stumbling block for the CPI; the 
    so-called clock ambiguity \cite{albrecht}.
    To elaborate, along with the requisite condition of strong entanglement,  
a  `good' clock in the CPI  should satisfy  two other requirements 
(see, {\em e.g.}, \cite{vedral}). 
The first is that the clock  should be able to  serve as an effective measuring device, 
meaning that it can access a
   sufficient amount of distinguishable states. The second is that the clock 
$C$ be weakly interacting
    with the remainder $R$, as interactions would naturally threaten the 
degree of their entanglement and also blur the delineation  of the two systems.    The latter condition for the clock is often extended to 
  the  limit of zero interactions, leading to the notion of
an `ideal' clock system. 
    The essence of the clock ambiguity problem is then  the existence of a  large (and possibly infinite) number of  choices for  good clocks, so that
any such description of $R$'s  dynamics is somewhat arbitrary.
 
Not too long ago,  Marletto and Vedral resolved this ambiguity by arguing that
it is natural to  limit considerations to ideal clocks and any 
choice of  ideal clock is related to any 
    other by a unitary transformation \cite{vedral}. 
On the other hand,  one might argue ---  as we recently did \cite{us} ---   
that interactions are an inevitable consequence of  a realistic  Universe and, as
such, cannot be dismissed out of hand or even taken to zero as  a limiting case.
Our previous investigation in \cite{us} considered a  semi-classical clock; 
the coherent-state description of a damped harmonic oscillator.~\footnote{Our working assumption in both \cite{us} and the current analysis is  that the size of the clock system  is small enough in comparison
to its complement   $R$ for the interactions to have a negligible effect on the
latter. For a different approach, see \cite{Smith}.}
  Following a procedure that was motivated in part by \cite{CC}, we found that the ideal-clock
limit  was not the optimal choice as far as it concerns   minimizing  uncertainty in
the clock readings. As it happens, this uncertainty {\em decreases} monotonically as the damping grows stronger. And so, given the previously discussed importance  of having  relatively weak interactions,  the optimal  choice for a damping
parameter is small but finite, and  it depends inversely on the running time of
the clock. It is implicit in this conclusion that the clock can only run
efficiently for a finite duration before a `resetting' is required.~\footnote {
See 
Subsection~\ref{1.2} for our actual  meaning of `resetting'.}  Otherwise, one could
simply impose the double scaling limit of infinitesimally weak  damping and
 an  infinitely long running time. This time limit is important in what follows.

    In order to advance  our investigation into the use of interacting clocks, we sought out a system
    with a truly quantum description.  Atomic clocks, as first suggested in practice by Rabi \cite{rabi}, fit quite naturally into this picture given that they are subjected to  decohering interactions. As will be made clear later 
in the paper, decohering atomic clocks are not much different than
damped, coherent oscillators with regard to  large uncertainties in the clock readings  being correlated
with weak interactions. What is different, however, is
the option of an infinitely long running time. This choice  is ruled out by fiat  in the current scenario, meaning that the double scaling limit is no longer
in play.

In spite of the small sample size for clock systems, we will further assert 
    that the incorporation of interactions is a generic requirement for
the CPI. 
  This argument is based upon exposing the properties of completely isolated systems.  In a similar manner to the above treatment of $\ket{\psi}$, the conserved total energy of an isolated system
    would restrict any description of its time evolution to the absolute time
parameter which is prescribed by the Schr{\"o}dinger equation. 
So that, in spite of previous claims to the contrary, 
 ideal clocks  can only provide a static, non-evolving description which is as
    timeless as the Universe in the Wheeler--De Witt equation. 
    \par
   
\subsection{`Disclaimer'}
\label{1.2}
Before proceeding, let us briefly comment on the perspective of the current paper and its authors. As  will be argued in an upcoming  discourse \cite{us2}, it is  `the problem with time' that is the real  problem and not time itself. Nevertheless, we would argue that, irrespective of any problem with time, the basic 
premise  of the Wheeler--DeWitt equation ---  that the Universe is inherently timeless --- must be correct even though the equation itself may well be flawed.
This stands to reason given that the Universe is a closed system; meaning that,
as it is employed here,  the Wheeler--DeWitt  equation (as well as 
the CPI by extension),   should be regarded as a metaphorical
or toy-model description of a more realistic and intricate  picture.  As such,  we would then also argue that the subsequent discussion is relevant regardless  of one's personal stand on either  the alleged  problem  or the  equation in question. With this as our current mindset, any discussion regarding the interpretation 
of time and the problems thereof will be kept to a minimum (however, see \cite{us2}).

It should be further noted  that the notion of a `resetting' time, which was introduced in
\cite{us} and motivated in analogy to ordinary timepieces, is not meant to imply  that some outside agent is needed to formally 
reinitiate the timing procedure. The time of resetting rather means that for which  the perturbative formalism
breaks down and, then, either the clock can  no longer effectively serve its purpose  or 
a more sophisticated treatment is required. And, because of the above 
viewpoint,  we are in no way suggesting  that temporal evolution  would, at this point,  
come to a crashing halt in the physical Universe. Nonetheless, a finite duration  for the Universe, if it is isolated, is not unreasonable insofar as it would
eventually have to  stop evolving on account of the second law of thermodynamics or its accelerated expansion (or both). 

It is important to keep in mind that the time $t$ of
the atomic clock  is not,  itself, the time which is  ``seen'' by   the remainder of the  Universe. This $t$ plays  the same role  as, for example,  Dolby's aforementioned abstract time \cite{dolby} or, in other words, it is simply an integration variable. In this version of the Page--Wootters method, the actual time parameter  would rather be the value of 
some  observable property of the clock system.   Provided that the clock system $C$ and its complement, the remainder  $R$, are maximally entangled, the states  of
$R$ would necessarily be correlated to the eigenstates of the relevant 
operator and, thus, with its eigenvalues as well. Meaning that the clock operator in question need not be the  Wheeler--DeWitt Hamiltonian (with the $R$ states 
traced out) as it is in more standard versions of  the Page--Wootters framework.  See 
Appendix~\ref{PWM} for further clarification on this methodology. 
For the case of an  atomic clock, in particular, $R$'s perceived time would  be related to
the inverse of the clock's resonant frequency.  Note, though, that  our current  interest is
with the efficiency of the clock rather than the actual clock readings.  

One final comment: For the discussion on isolated systems in Section~4,
the arguments apply just as well to classical (sub)systems as they do to 
quantum ones.

\subsection{Contents}
\label{1.3}
 The remainder of the paper is laid out as follows. 
    The next section briefly describes the atomic-clock procedure and its applicability to the CPI.
    Section~\ref{3}  reports on the effects of decoherence and identifies 
the optimal atomic clock from the CPI perspective.
    Section~\ref{4} presents a general argument for our claim that interactions are a necessary feature in any consistent 
    description of time evolution.
    Section~\ref{5} provides a brief summary, and 
 some additional details about the maths 
  are included  in four appendices.

  \section{The atomic clock}
  \label{2}
    
    In 1945, Rabi  presented the first practical approach for obtaining a time measurement using atomic frequencies \cite{rabi}.
    (See, {\em e.g.}, \cite{sak} for a textbook account.)
    The method provides a time measurement by counting the cycles of an electromagnetic oscillator and dividing by 
    the oscillator frequency $\omega$. 
 A standardized  unit of time can be defined by   setting the frequency $\omega$ to the transition frequency of an electron 
   in a particular atom. 
We will adopt the notation 
     $\;\omega_{21}=E_2-E_1\;$ for the transition frequency, where $E_{1,2}$ are the ground and excited state respectively and $\hbar$ has been set to unity
    here and throughout. In order to ensure that $\omega$ is as close to the desired 
    transition frequency as possible, the atoms in question are set in the ground state and then
    exposed to the oscillator. 
    By modulating $\omega$, one  will change the probability  of finding the 
exposed electrons in the excited state  and can then  plot this probability $P_{ex}$ versus $\omega$.
    The maximum value for $P_{ex}$ corresponds to the resonant frequency,
 $\; \theta\equiv\omega-\omega_{21}=0\;$,
    and the
 full
    width at half maximum (FWHM) of the plot measures $\delta\omega$, the uncertainty in $\omega$ (and, consequently, that of the  time measurements). 
One finds that 
$\;\delta\omega\propto\lambda\;$, where $\lambda$ is the amplitude of the oscillating wave.
 
    Improvements to the Rabi method were later made by Ramsey \cite{ramsey}.
That author showed that exposing the atoms to the oscillator 
    for two short times (or pulses) $\tau$, separated by a longer non-interaction time $T$, would reduce the 
    uncertainty of the measurements \cite{ramsey}. 
    When  the value of the frequency is sufficiently close to  resonance, $\;\theta\ll\lambda\;$,  its uncertainty  rather  goes as $\;\delta\omega=\frac{\pi}{T}\;$, where $T$ has become known as the Ramsey time.
    This suggests that taking the limit $\;T\rightarrow\infty\;$ would minimize the uncertainty. However, $T$ 
    must indeed be finite as the Ramsey process still requires a period of
exposure  $\tau$ to take place immediately {\em after} $T$.

  The atomic clock could, of course,  be made arbitrarily accurate by setting the system 
    to the resonance case.  Assuming, for the sake of argument, that such accuracy could be achieved at least   as a matter of principle, one ends up with  a description of time that cannot be distinguished from 
    the absolute time parameter  already appearing in the Schr{\"o}dinger equation.  Ultimately, we will claim that the limiting case of a non-interacting clock 
 leads to a time description 
    which is similarly indistinguishable from absolute time, 
    as  the CPI then fails  to account for our  passage through time. 
      To prove this, the atomic clock will be allowed to interact
with the rest of the Universe through the inclusion of  decoherence effects.
      
  \section{The decohering atomic clock}
  \label{3}
  
    In order to analyze the effects of decoherence on the atomic clock,
we will be incorporating the dynamics of the Linblad equation \cite{lind}.
The basic idea is to  allow the clock system to decohere during the Ramsey 
interval $T$, as the time $\tau$ of the oscillator pulse is taken
to  be small enough to ignore the effects of decoherence during these brief periods of exposure. A significant portion  of our method follows an approach that was sketched out by Weinberg \cite{Wein}.

As outlined in Appendix~\ref{A}, the first step is to derive  the  evolution operator for the
Ramsey setup when decoherence is included. The  next step is to use this
operator to calculate  the probability  of finding the system in 
    the excited state $P_{ex}(t)$ after both pulses and the Ramsey time have transpired. 
As explained in Appendix~\ref{B}, this process leads to  
    \begin{equation}
     \label{P_full}
     \begin{split}
     P_{ex}(2\tau+T)&=
     \frac{4\lambda^2}{\Omega^2}\sin^2(\Omega\tau)\Big[2\cos(\Omega\tau)+\frac{\theta^2}{2\Omega^2}\sin^2(\Omega\tau)\\
     &+e^{-\alpha T}\Big(2\cos^2(\Omega\tau)\cos((\theta-\beta)T)\\
     &-\frac{\theta}{2\Omega^2}\sin^2(\Omega\tau)\cos((\theta-\beta)T)\\
     &-\frac{2\theta}{4\Omega}\sin(2\Omega\tau)\sin((\theta-\beta)T)\Big)\Big]\;,
	\end{split}
    \end{equation}
    where $\alpha$ and $\beta$ are the real and imaginary parts of the `decoherence factor' $\gamma$ ({\em i.e.}, $\gamma$ is  one of the eigenvalues of the non-unitary portion
of the Linblad equation) and
    $\;\Omega=\sqrt{\lambda^2+\frac{\theta^2}{4}}\;$ is known as the Rabi frequency.  

Following Ramsey, we will fix  the  pulse time $\tau$  by maximizing the probability for the idealized case of $\;\theta=T=0\;$ \cite{ramsey}. Making this choice and setting $\;P_{ex}(2\tau+T)=1\;$, one finds  that  $\;\tau=\frac{\pi}{4\lambda}\;$.
 Then, with the
    substitution of $\tau$ and the  assumption that $\;\theta\ll\lambda\;$ ({\em i.e.}, the system is close to resonance), eq.~(\ref{P_full}) reduces down to
    \begin{equation}
     \label{P}
      P_{ex}(2\tau+T)\;=\;\frac{1}{2}\Big[1-e^{-\alpha T}\cos((\theta-\beta)T)\Big]\;.
    \end{equation}
    This expression closely resembles one  from \cite{Wein}, where it was 
applied in a 
    different context.  
    
The uncertainty in $\omega$ for the current case ---  again  calculated as the FWHM from the plot of $P_{ex}$ vs $\omega$ --- is
    found to be  $\;\delta=\frac{\pi}{T}\;$, exactly the same as before. 
   This would suggest  that the inclusion of  decoherence has no bearing on the precision of the measurements. However, this is not true because 
    the maximum outcome for the probability has definitely diminished. 
 Put into more physical terms, the decoherence of the system reduces 
its  ability to function as a quantum clock. More rigorously, a  constrained minimization of the uncertainty in the probability 
$P_{ex}(2\tau+T)$ leads to the following relation, 
 \begin{equation}
     \label{at}
      \alpha T\;\sim\;\mathcal{O}(1)\;,
\end{equation}
as elaborated on in Appendix~\ref{C}.

   The above outcome indicates that 
the desire  for a long Ramsey time $T$ (which must anyways  be finite) for the purposes of 
    minimizing the uncertainty  must be 
    balanced against the (similarly finite) effects of decoherence. This is really just another way
of justifying the previously stipulated  condition  of a weakly interacting clock, which 
translates into
 $\;\alpha T\lesssim\mathcal{O}(1)\;$.
    
And so, in attempting to impose the ideal-clock limit of 
$\;\alpha\rightarrow0$\;, one is stymied by both 
    the condition of a finite $T$ and the proclivity for more accurate measurements.
    Our conclusion is that the ideal limit of an  atomic clock 
    is neither a tenable nor an optimal choice. 
    
  \section{What can be said about ideal clocks}
  \label{4}
  
    The story that the above result seems to be telling is one where interactions are a necessary feature
    in the framework of the CPI. The current objective is
to both generalize and strengthen 
   our conclusions about  atomic clocks (and, previously, coherent states \cite{us}). This will be accomplished with an independent, qualitative argument.
 
    Let us start by reconsidering the Wheeler--DeWitt equation. Its timelessness  can be attributed to the Universe 
    having a total net energy of zero, as per the right-hand side of eq.~(\ref{wdw}).
    The precise value of the energy, however, is  really besides the point.  Any (strictly) constant value for the 
    energy would imply that the dynamics of the Universe are frozen, rendering time a meaningless concept. 
   Let us now  consider the more familiar case of an  isolated (sub-)system 
as it would be described in a textbook on  quantum mechanics. The dynamics
of this system are 
    similarly frozen, yet we attribute it with a time parameter all the same; namely that of the time-independent 
    Schr{\"o}dinger equation. Where did this time come from? There are only two possibilities: the system's 
    notion of time was put in by hand or it was inherited from a larger, ancestral system. 
   But can the  idealized clock system of the CPI   answer this same question 
about the the origin of {\em its}  notion 
    of time? The CPI `rulebook'  does not permit us to put in time by hand and neither can the clock inherit its time from 
    an ancestor, as the only one available is the timeless Universe {\em \`a la}
 Wheeler and DeWitt.
    And so, with no notion of time available and no opportunity to interact with its environment, the ideal clock
    cannot possibly evolve relative to another system or component thereof. In short, the idealized clock 
    could never serve as a timepiece for another system any more or less than
the Universe as a whole could.

    A sequence of states could  still be described for these timeless, isolated  systems as
 illustrated in
    Marletto and Vedral's treatment \cite{vedral}. Each successive state of $R$ is identified with 
    a time measurement of $C$ and a history is produced. However, as pointed out by those same authors, this picture provides neither 
    a flow of time nor an arrow of time ---  both of these concepts should be 
viewed as  fictitious within this timeless framework.
    There is simply no motivation for moving from one state to the next, and no 
provision for
  a sense  of movement through time without also assuming an 
absolute, external time 
   along  with an imposed direction. 
    The ideal-clock scenario then precludes the possibility of a clock which itself can  experience time or can provide a 
    measurement of  time for an external agent.
    \par
    Taking our lead from the above argument and including interactions as a matter of principle, we arrive at 
    a very different result. 
    The requirement of an open system for $C$ immediately allows for a 
 clock with a sense of 
  evolving in  time and, likewise, for its complement $R$. 
    The resulting time evolution includes a description 
   of  not just the history of states for $C$ and $R$ but also  
an arrow in time thanks to the
    non-reversible effects of decoherence and/or damping.
    We thus have 
    a way of reconciling our experience of passing through time with the timeless state of the  Universe. 
          
   \section{Conclusions}
   \label{5}
    
    Our investigation into atomic-clock systems showed 
    that the optimal choice of clock requires a compromise between  
fending off the effects of decohering 
    interactions and  maximizing the accuracy of the clock. This reinforced a previous result on coherent-state clocks 
    and led to a new view on the description of time within the framework of
the CPI. 
    The restriction to the ideal-clock (non-interacting) limit prohibits any description of motion through time;
  there can   only be a static series of states with no motivation for
any movement between them. 
    The inclusion of interactions, however, resolves this issue as the interacting clock system can 
    evolve through its relation with the complementary system. 
Elevating this framework  to  `reality' (or, rather, some  simplistic description thereof), 
   one would  translate this evolution into a passage through time, rather than the inclusion of an abstract, absolute time dimension 
    which sits `outside' of our experience. 
    Our conclusion is that the use of an ideal clock in the CPI not only 
fails at being  the optimal choice in practice but also 
   represents a misleading assumption in principle; interactions must be included as a strict rule. 
    \par
    This motivation to use only interacting clocks within the CPI does, however, reintroduce the
    clock ambiguity as the ideal-clock limit can no longer  be called upon to resolve the issue. 
    As the inclusion of interactions  appears to come along  with a free arrow of time,  
   there could well  be a  solution to the clock ambiguity which utilizes
    a  preference for clocks obeying the second law, rather than appealing to 
the redundancy of ideal clocks. 
    This possibility  is 
    currently under investigation \cite{us2}.

 \section*{Acknowledgments}
          The research of AJMM received support from an NRF Incentive Funding Grant 85353 and  NRF Competitive Programme Grant 93595.  KLHB is supported by an NRF bursary through  Competitive Programme Grant 93595 and a Henderson Scholarship from Rhodes University. This work is based on the research also supported in part by the National Research
              Foundation of South Africa (Grant Numbers: 111616).
    
   \appendix

\section{Evolution according to Page and Wootters}
\label{PWM}

Section \ref{1.1} outlined the Page--Wootters method of recovering time. The method is based on the timeless description of the Universe as a pure state $\ket{\psi}$, which is governed by the 
Hamiltonian $\hat{H}$ in the Wheeler--DeWitt equation. 
Here, we describe the method in more detail along with an explanation of the role of the abstract variable $t$ which is discussed in Section \ref{1.2}.

The standard description involves the division of $\ket{\psi}$ into the clock $C$ and the rest $R$. This partitioning is accompanied by the identification of the Hamiltonians
$\;\hat{H}_C={\rm Tr}_R\hat{H}\;$ and $\;\hat{H}_R={\rm Tr}_C\hat{H}$, which govern $C$ and $R$ respectively.

Interaction effects (governed by $\hat{H_I}$) between $C$ and $R$ complete the Hamiltonian, which can be written as $\hat{H}=\hat{H}_C\otimes\mathbbm{1}+\mathbbm{1}\otimes\hat{H}_R+\hat{H}_I\;$.
Under the Page--Wootters method, these interaction effects are considered vanishingly weak and so $\hat{H}_I$ can be ignored. This leads to  the approximate relation  
 $\;\hat{H}\approx\hat{H}_C\otimes\mathbbm{1}+\mathbbm{1}\otimes\hat{H}_R\;$
and, because $\;\hat{H}|\psi\rangle=0\;$ for physical states, it follows that 
\begin{equation}
\label{h}
\hat{H}_C\;\approx\;-\hat{H}_R\;
\end{equation}
is true when acting on physical states.

The last requirement for the Page--Wootters method is that $C$ and $R$ be in a
maximally  entangled state, 
\begin{equation}
 \label{ent}
 \ket{\psi}\;=\;\sum_j \alpha_j \ket{\psi_C}_j\ket{\psi_R}_j\;,
\end{equation}
where $\ket{\psi_{C,R}}$ are states for  $C$ and $R$ respectively, a subscript of $j$ indicates a basis state and $\alpha_j$ represents numerical coefficients. 
In this way, the evolution of $C$ can be `transfered' to $R$,
\begin{equation}
\begin{split}
  \ket{\psi_j}\;=\;c_j\left(e^{-i\hat{H}_Cp}\ket{\psi_C}_j\right)\ket{\psi_R}_j
             \;=\;&c_je^{-i(\hat{H}-\hat{H}_R)p}\ket{\psi_C}_j\ket{\psi_R}_j\\
             \;\approx \;&c_j\ket{\psi_C}_j\left(e^{i\hat{H}_Rp}\ket{\psi_R}_j
\right)\;,
 \end{split}
\end{equation}
where $p$ refers to the eigenvalues of the conjugate to $\hat{H}_C$; in other
words, $p$ is the emergent time parameter for $R$. Note that we have set
$\;\hbar=1\;$, here and throughout.

Up to this point, the standard description of the Page--Wootters method is sufficient. But, in order to analyze the case of the  atomic clock, we use
a variant that was inspired by Dolby \cite{CC}.
What is now needed is some observable property of $C$ (but not necessarily
$p$) to act as
the time parameter for $R$. Let us denote this property by  $x$. (In \cite{us}, $x$
was literally a position variable. For the atomic clock, $x$ would be
related to the inverse of the resonant frequency.)
Let us further denote the conjugate to the operator that measures $x$ as
$\hat{\Phi}_C^x$. Then the condition of maximal entanglement is enough
to ensure that there is a ``mirror'' operator  acting on states of $R$,
$\hat{\Phi}_R^x$,
for which
\begin{equation}
\hat{\Phi}_C^x\;\approx\;-\hat{\Phi}^x_R\;
\end{equation}
is true when acting on physical states and for some suitable choice
of bases. Meaning that $\hat{\Phi}_{C,R}^x$ can (and do) play the role of
effective Hamiltonians.

Let us now consider how the abstract time parameter $t$ fits in and understand
why it does not require a physical interpretation.
If $t$ is the `conventional' time parameter for the clock operator,
 there should be some semiclassical relation $\;x=x(t)\;$.
Then the probability for the clock to be in  a state for which  $\;x=x'\;$ 
can be expressed  as an integral over the probability
that $\;x=x'\;$ when conditioned on $\;t=t'\;$.  In other words,
\begin{equation}
 \label{prob2}
 P_C(x')\;=\;
 \int^{\infty}_{-\infty} dt'\; |\langle x| \psi_C(t')\rangle|^2\;.
\end{equation}
And so
$t$ is merely an intergation variable as advertised.
Moreover, because of the condition of maximal entanglement, the relationship
\begin{equation}
P_R(x')\;\approx\;P_C(x')
\end{equation}
immediately follows.

    \section{Determining the evolution operator}
    \label{A}
      
      Here, Ramsey's evolution matrix \cite{rabi} is generalized
 to allow for decoherence. 
      \par
   Let us consider a two-level system at time $\;t=0\;$  in its  energy basis,
 $\;\ket{\psi(0)}=c_1\ket{1}+c_2\ket{2}\;$. 
      If the external potential oscillates according to 
      \;$V(t)=\lambda e^{i\omega t}+\lambda e^{-i\omega t}\;$  with $\lambda$ real, then the state at a later time  $t$ is    
      $\;\ket{\psi(t)}=\Big(\cos(\Omega t)-\frac{i\theta}{2\Omega}\sin(\Omega t)\Big)e^{i\theta t/2}\ket{1}
	         +\frac{\gamma e^{-i\theta t/2}}{i\Omega}\sin(\Omega t)\ket{2}\;$
      and its density matrix is
      \begin{equation}
       \label{A1}
       \rho(t)=\ket{\psi(t)}\bra{\psi(t)}=\begin{bmatrix}1-\frac{\gamma^2}{\Omega^2}a^2 &
		    \frac{i\gamma e^{i\theta t}}{\Omega}a\Big(b-\frac{i\theta}{2\Omega}a\Big) \\
		    -\frac{i\gamma e^{-i\theta t}}{\Omega}a\Big(b+\frac{i\theta}{2\Omega}a\Big) 
		    & \frac{\gamma^2}{\Omega^2}a^2\end{bmatrix},
      \end{equation}
      where $\;a=\sin(\Omega t)\;$, $\;b=\cos(\Omega t)\;$  and $\;\Omega=\sqrt{\lambda^2+\frac{\theta^2}{4}}\;$.
 Alternatively, the evolution can be described by $\;\rho(t)=U(t,0)\rho(0)U^\dagger(t,0)\;$.

      Assuming that the system  starts in its ground state,  $\;\rho(0)=\begin{bmatrix}1&0\\0&0\end{bmatrix}\;$, one 
    find an evolution matrix of the form
      \begin{equation}
       \begin{bmatrix}\rho(t)_{11}&\rho(t)_{12}\\\rho(t)_{21}&\rho(t)_{22}\end{bmatrix} \; =\;
       \begin{bmatrix} A&B\\-B^*&A^*\end{bmatrix}
       \begin{bmatrix} 1&0\\0&0\end{bmatrix}
       \begin{bmatrix} A^*&-B\\B^*&A\end{bmatrix}\;.
      \end{equation}
      where
      \begin{equation}
       \label{A3}
       \begin{split}
       |A|^2\;=\;&b^2+\frac{\theta^2}{2\Omega^2}a^2\; ,\\
       |B|^2\;=\;&\frac{\lambda^2}{\Omega^2}a^2\; ,\\
       -AB\;=\; &\frac{i\lambda e^{i\theta t}a}{\Omega}(b-\frac{i\theta}{2\Omega}a)\;,\\
       -A^*B^*\; = \;&\frac{-i\lambda e^{-i\theta t}a}{\Omega}(b+\frac{i\theta}{2\Omega}a)\;.\\
       \end{split}
      \end{equation}
      Since $\;A=b-\frac{i\theta}{2\Omega}a\;$ is true up to a phase, the evolution matrix can be resolved into
      \begin{equation}
       \label{A4}
       U(t,0)\; = \;\begin{bmatrix}
               b-\frac{i\theta}{2\Omega}a & -\frac{i\lambda e^{i\theta t}}{\omega}a\\
               -\frac{i\lambda e^{-i\theta t}}{\omega}a & b+\frac{i\theta}{2\Omega}a
              \end{bmatrix} \;.
      \end{equation}
      The non-decohering limit of this outcome agrees with the final form of Ramsey's result  \cite{ramsey}.

    \section{Calculating the probability}
    \label{B}
      
      The following is a calculation of the probability of finding the system in its excited state
      after the Ramsey procedure has been completed.
      \par
      The first  pulse (or exposure zone) changes the system from the ground state according to 
      $\;\rho(\tau)=U(\tau,0)\rho(0)U^\dagger(\tau,0)\;$.
      To include decoherence  during the Ramsey time period $T$, 
the following  Lindblad equation \cite{lind} is applied \cite{Wein}:
      \begin{equation}
       \label{B1}
       \dot{\rho}\;=\;-i[\hat{H},\rho(t)]+\sum_\alpha\Big[\hat{L}_\alpha\rho(t)\hat{L}^\dagger_\alpha
		      -\frac{1}{2}\hat{L}^\dagger_\alpha\hat{L}_\alpha\rho(t)
		      -\frac{1}{2}\rho(t)\hat{L}^\dagger_\alpha\hat{L}_\alpha\Big]\;.
      \end{equation}
      The effect of this evolution on the system is 
      \begin{equation}
      \label{B2}
	\rho(t)_{mn}\;\propto\; e^{-i(E_m-E_n)t-\gamma_{mn}t}\;=\;e^{-i(E_m-E_n)t-\gamma_{mn}t}\;,
      \end{equation}
      where $m,\;n$ label the eigenstates of the operators on the right-hand side of eq.~(\ref{B1})
      and     
$\gamma_{mn}\;=\;\alpha+i\beta=(\alpha-i\beta)^*=\gamma_{nm}^*\;$ represents the decoherent part of their eigenvalues (the state labels on $\alpha$ and $\beta$ are implied).

It can be seen that the diagonal terms of $\rho(\tau)$ are insensitive to the decoherence. On the other hand, 
      the effect on the off-diagonal terms
      is evident from
      \begin{equation}
       \label{B3}
	\rho(\tau+T)\;=\;\begin{bmatrix}
b^2+\frac{\theta}{4\Omega^2}a^2 &\frac{i\lambda e^{i\theta}e^{(i\omega_{21}-\gamma)T}}{\Omega}a(b-\frac{i\theta}{2\Omega}a)\\
\frac{-i\lambda e^{-i\theta}e^{(-i\omega_{21}-\gamma^*)T}}{\Omega}a(b-\frac{i\theta}{2\Omega}a)& b^2+\frac{\theta}{4\Omega^2}a^2
	             \end{bmatrix}\;,
      \end{equation}
      where $a$ and $b$ have been defined after eq.~(\ref{A1}).
      
It should  be noted that the external potential $V(t)$ continues to oscillate for the 
      duration of $T$. This introduces an additional  phase factor $e^{\pm i\omega T}$ in the  
      off-diagonal terms of the evolution operator once the second pulse 
$\tau$ is applied.
      The equation governing this last exposure zone is given by
     \begin{equation}
       \label{B4}
       \begin{split}
	&\begin{bmatrix}\rho(2\tau+T)_{11}&\rho(2\tau+T)_{12}\\\rho(2\tau+T)_{21}&\rho(2\tau+T)_{22}\end{bmatrix}\; =\;\\
	&\begin{bmatrix} A&B\\-B^*&A^*\end{bmatrix}
	\begin{bmatrix}\rho(\tau+T)_{11}&\rho(\tau+T)_{12}\\\rho(\tau+T)_{21}&\rho(\tau+T)_{22}\end{bmatrix}
	\begin{bmatrix} A^*&-B\\B^*&A\end{bmatrix}\;,
       \end{split}
      \end{equation}
      where $A$ and $B$ have been defined in eq.~(\ref{A3}).
      
The element $\rho(2\tau+T)_{22}$ represents the probability of finding the system in its excited state $P_{ex}(2\tau+T)$  and is determined to be 
      \begin{equation}
       \label{B5}
       \begin{split}
 P_{ex}(2\tau+T)\; = \; \rho(2\tau+T)_{22}\;=\;&\frac{4\lambda^2}{\Omega^2}\sin^2(\Omega\tau)\Big[2\cos(\Omega\tau)+\frac{\theta^2}{2\Omega^2}\sin^2(\Omega\tau)\\
     &+e^{-\alpha T}\Big(2\cos^2(\Omega\tau)\cos((\theta-\beta)T)\\
     &-\frac{\theta}{2\Omega^2}\sin^2(\Omega\tau)\cos((\theta-\beta)T)\\
     &-\frac{2\theta}{4\Omega}\sin(2\Omega\tau)\sin((\theta-\beta)T)\Big)\Big]\;,
       \end{split}
      \end{equation}
      which correctly reduces to Ramsey's result when $\;\alpha=\beta=0\;$.
    
    \section{Taking limit of uncertainty}
    \label{C}

     The goal here is to use a more rigorous method to  substantiate 
the claim in Section~\ref{3} that $\;\alpha T\sim\mathcal{O}(1)\; $.

 To quantify the  effect of decoherence on a clock measurement, we will calculate
and then minimize the relative uncertainty of     
    of $P_{ex}(2\tau+T)\;$; that is,  $\;\frac{\delta P_{ex}}{P_{ex}}\;$ (with the time dependence now left implicit). To start, let us consider the relation
    $\;\delta P_{ex}=\sqrt{\big(\frac{\partial P_{ex}}{\partial\omega}\big)^2\delta\omega^2}\;$, which then gives 
    \begin{equation}
     \label{P_unc}
    \frac{\delta P_{ex}}{P_{ex}}\;=\; \frac{\pi}{2}e^{-\alpha T}\sin(\Theta T)\Big(1+e^{-\alpha T}\cos(\Theta T)\Big)^{-1}\;,
    \end{equation}
    where the definition  $\;\Theta=\theta-\beta\;$ has been applied.

Eq.~(\ref{P_unc}) makes  it clear that the limit  $\;T\rightarrow\infty\;$ minimizes the uncertainty. 
    However, the restriction on a finite value for $T$ must still be in place in order to complete the 
    procedure as prescribed by Ramsey. Given that the constraint (see Section~2) $\;T\sim\frac{1}{\delta\omega}\;$ is also in effect --- 
    which also  ensures a finite  $T$ barring the classical limit ---  the  minimization procedure then amounts to
      solving      
      \begin{equation}
       \label{C1}
       \frac{\partial}{\partial T}\Bigg[\frac{\delta P_{ex}}{P_{ex}}-\Lambda(\delta\omega-\frac{\pi}{T})\Bigg]\;=\;0\;,
     \end{equation}
      where $\Lambda$ is a Lagrange multiplier.
     With the help of eq.~(\ref{P_unc}), the above expression  resolves into 
      \begin{equation}
       \label{C2}
       e^{-2\alpha T}\Big(\Theta T^2 -2\Lambda\cos^2(\Theta T)\Big)+e^{-\alpha T}\Big((\Theta T^2-4\Lambda)\cos(\Theta T)
  -\alpha T^2\sin(\Theta T)\Big)-2\Lambda\;=\;0\;.
      \end{equation}
      
Looking at $\;\Theta=\theta-\beta\;$ and knowing that $\theta$ will vanish as the resonant case is approached, we can apply the 
       approximation $\;|\Theta|\approx-\beta\;$.
      We can further approximate $\;\beta\sim\pm\alpha\;$, as $\alpha$ is essentially 
      the only dimensional scale in the problem.
      With these simplifications, it can now  be readily checked that  either of the limits $\; \alpha T\gg1\; $ or 
$\; \alpha T\ll1\; $  implies
      that $\; \Lambda=0\; $. However, one cannot argue that $\Lambda$ vanishes as this choice effectively
      removes the finiteness condition on $T$. 
      That leaves  $\;\alpha T\sim\mathcal{O}(1)\;$ as the only viable solution.
      \par
      Even though this is not an explicit calculation, the approximations still allow us to 
      make a statement about the relationship between $\alpha$ and $T$. Specifically, the fact that 
      the two are related through a finite-valued product indicates that only one of the pair can be set independently.
      This is exactly why the finiteness of $T$ imposes the same condition on $\alpha$.

\end{document}